\documentclass[twocolumn,amsmath,amssymb,longbibliography,aps,prb]{revtex4-2}
\usepackage[dvipdfmx]{graphicx}

\usepackage{bm}
\usepackage{color}
\usepackage{braket}
\usepackage{lipsum}
\usepackage{mathdots}
\usepackage[version=3]{mhchem} 

\usepackage{xspace}
\usepackage[unicode=true,colorlinks=true,linkcolor=blue,citecolor=blue,urlcolor=blue]{hyperref}

\newcommand{\TMA}     {\ce{Ti2MnAl}\xspace}

\newcommand  {\eqn}[1]{(\ref{eqn:#1})} 
\renewcommand{\(}     {\left(}
\renewcommand{\)}     {\right)}

\renewcommand{\_}[1]  {_\textrm{#1}}

\begin{document}


\title{
Effective Tight-Binding Model of Compensated Ferrimagnetic Weyl Semimetal with Spontaneous Orbital Magnetization 
}

\newcommand{\authA}{\author{Tomonari Meguro$^1$}\thanks{meguro.tomonari@mbp.phys.phys.kyushu-u.ac.jp}}
\newcommand{\authB}{\author{Akihiro Ozawa$^{2}$}\thanks{Present address: Institute for Solid State Physics, University of Tokyo, Kashiwa 277-8581, Japan}\thanks{akihiroozawa@issp.u-tokyo.ac.jp}}
\newcommand{\authC}{\author{Koji Kobayashi$^{1}$}\thanks{Present address: Physics Division, Sophia University, Chiyoda-ku, Tokyo 102-8554, Japan}\thanks{k-koji@sophia.ac.jp}}
\newcommand{\authD}{\author{Kentaro Nomura$^1$}\thanks{nomura.kentaro@phys.kyushu-u.ac.jp}}

\newcommand{\affiA}{$^1$Department of Physics, Kyushu University, Fukuoka 819-0395, Japan}
\newcommand{\affiB}{$^2$Institute for Materials Research, Tohoku University, Sendai 980-8577, Japan}

 \authA
 \authB
 \authC
 \authD
 \affiliation{\affiA}
 \affiliation{\affiB}

\newcommand{\abstbody}{
The effective tight-binding model with compensated ferrimagnetic inverse-Heusler lattice \TMA, candidate material of magnetic Weyl semimetal, is proposed.
The energy spectrum near the Fermi level, the configurations of the Weyl points, and the anomalous Hall conductivity are calculated.
We found that the orbital magnetization is finite, while the total spin magnetization vanishes, at the energy of the Weyl points.
The magnetic moments at each site are correlated with the orbital magnetization, and can be controlled by the external magnetic field.

}
 \begin{abstract}
  \abstbody
 \end{abstract}

\maketitle

\section{Introduction}\indent
Topological semimetals are new classes of materials characterized by the topologically-nontrivial gapless nodes.
One of the representative systems is the Weyl semimetal~(WSM), 
a gapless semiconductor with non-degenerate point nodes called Weyl points~\cite{murakami2007,Wan2011,Burkov2011}. 
In momentum space, these nodes behave as a source or sink of a fictitious magnetic field~(Berry curvature), 
which is distinguished by the sign degrees of freedom, so-called chirality~\cite{xiao2010,Armitage2018}.
The Weyl points with the opposite chirality must appear in pairs~\cite{Nielsen1981,Nielsen1983}. 
Originating from the Weyl points, 
distinctive magnetoelectric effects, such as 
the chiral magnetic effect~\cite{Zyuzin2012,Chen2013,Vazifeh2013,Triangle2014},
arise. 
Generally, to realize the WSM, one has to break either the inversion or time-reversal symmetry of Dirac semimetal~\cite{Young2012,Liu2014stable,Liu2014}, which has degenerate gapless linear dispersions.
As early theoretical works, the WSM phases in an antiferromagnetic pyrochlore structure~\cite{Wan2011} and a topological insulator multilayer~\cite{Burkov2011} were proposed.
Especially, in Ref.~\onlinecite{Burkov2011}, Burkov et al. discussed the anomalous Hall effect~(AHE) in the WSMs with broken time-reversal symmetry, so-called magnetic Weyl semimetals~(MWSMs).
It was shown that 
the anomalous Hall conductivity~(AHC) is proportional to the distance between the pair of Weyl points with the opposite chirality.


After these theoretical predictions, exploring for the WSM phases in specific materials has been demonstrated.
In the early stage of the experimental studies, non-magnetic WSMs with the broken inversion symmetry, 
such as \ce{TaAs}, were examined~\cite{Xu2015_1,Xu2015_2}.
On the other hand, a great deal of effort has been devoted to realize the MWSMs.
After then, both theoretical and experimental studies have succeeded in discovering the MWSM phases in some systems, 
such as layered-kagome~\cite{Barros2014,Chen2014,Kubler2014,Nakatsuji2015, Kuroda2017,Suzuki2017,Balents2017,Ito2017,Ye2018,Liu2018,Xu2018,Wang2018,Liu2019,Ozawa2019,Takagaki2019,Kim2019,Shen2020,Guguchia2020,Tanaka2020,Thakur2020,Ikeda2021,Yanagi2021,Watanabe2022,Ozawa2022}   
and Heusler systems~\cite{Sakai2018,Reichlova2018,Guin2019,Webster1971,Kubler2016,Li2020,Umetsu2008}. 
The layered-kagome systems attract much attention from viewpoints of anomalous transport phenomena
~\cite{Chen2014,Kubler2014,Nakatsuji2015, Kuroda2017,Suzuki2017,Balents2017,Ito2017,Ye2018,Liu2018,Takagaki2019,Kim2019,Yanagi2021} 
and various magnetic orderings~\cite{Barros2014,Guguchia2020,Thakur2020,Watanabe2022,Ozawa2022}.
Antiferromagnetic \ce{Mn3Sn} shows the AHE even without net magnetization~\cite{Chen2014,Kubler2014,Nakatsuji2015, Kuroda2017,Suzuki2017,Balents2017,Ito2017}.
Ferromagnetic \ce{Co3Sn2S2} shows the giant AHE and small longitudinal conductivity,
resulting in
the large anomalous Hall angle reaching 20$\%$~\cite{Liu2018,Xu2018,Wang2018,Liu2019,Ozawa2019,Shen2020,Tanaka2020,Ikeda2021,Watanabe2022}.
As other promising candidates, recent studies reported ferromagnetic Heusler systems with relatively high Curie temperature $T_{\rm C}$ compared to those of the layered-kagome systems.
For example, \ce{Co2MnGa}~\cite{Sakai2018,Reichlova2018,Guin2019} with $T_{\rm C}\approx$ 694~K~\cite{Webster1971} and 
\ce{Co2MnAl}~\cite{Kubler2016,Li2020} with $T_{\rm C}\approx$ 724~K~\cite{Umetsu2008} are also studied.
In addition to the layered-kagome and Heusler systems, 
other systems such as \ce{EuCd2As2}~\cite{Ma2019,Wang2019,Soh2019} 
and $Ln$Al$Pn$~\cite{Xu2017,Chang2018} ($Ln=$ lanthanides, $Pn=$ Ge, Si) 
are also reported as candidate materials of MWSMs.
A wide variety of candidates for MWSMs has been explored and has influenced on both the field of topological materials and magnetism.
The magnetic Weyl semimetal phase has been reported in compensated ferrimagnetic inverse Heusler alloy \TMA~\cite{Shi2018} by first-principles calculations.
In \TMA, 
magnetic moments at two Ti sites are anti-parallel to that at the Mn site, showing zero net magnetization. 
The transition temperature is determined to be 650 {\rm K}~\cite{feng2015z}, which is comparable to those of other Heusler Weyl systems, such as \ce{Co2MnGa}.
%
%
%
Besides, 
\TMA exhibits a large AHE despite its vanishing total spin magnetization, similar to \ce{Mn3Sn}.
However, in contrast to \ce{Mn3Sn}, the density of states at the Weyl points in \TMA is negligibly small, 
resulting in the manifestation of a large anomalous Hall angle compared to other MWSMs.
These properties indicate that 
the unique electronic and magnetic structures of \TMA 
provide 
distinctive magnetoelectric response and spintronic functionalities compared to conventional materials.

In order to study the magnetoelectric responses specific to \TMA, a quantitative analysis is necessary.
First-principles calculations are widely used to precisely calculate the electronic structure, considering all the electron orbitals.
However, with the first-principles calculations, 
it is generally difficult to study the magnetoelectric responses related to the complicated spin texture, such as the magnetic domain wall.
This is because the matrix of the Hamiltonian becomes huge due to the lack of translational symmetry.
%
%
Therefore, a simple tight-binding model describing the Weyl points of \TMA is beneficial to calculate these magnetoelectric responses.

In this paper, 
we construct an effective tight-binding model of \TMA using a few orbitals.
We consider the single orbitals of two Ti and Mn, spin-orbit coupling, and exchange interaction between compensated ferrimagnetic ordering and itinerant electron spin.
Using our model,
we study the electronic structure, AHE, spin and orbital magnetizations, and magnetic anisotropy.
We show that our model describes the configuration of the Weyl points that is consistent with the results obtained by the first-principles calculations.
Also, we discuss the control of compensated ferrimagnetic ordering by orbital magnetization.



\section{Model}
\begin{figure}[tb]
   \centering
   \includegraphics[width=1.0\hsize]{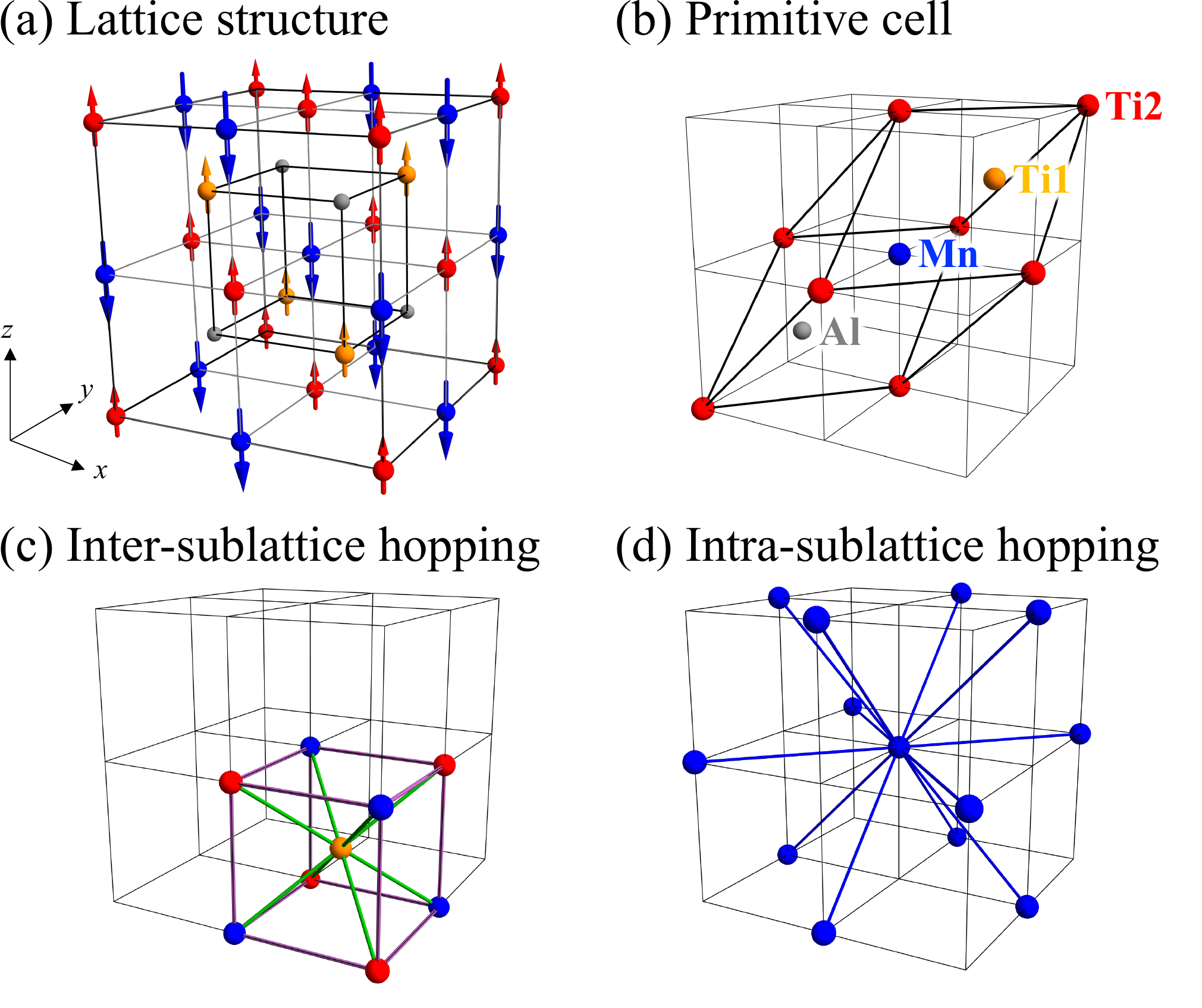}
\caption{
(a) Conventional unit cell of \TMA.
Ti and Mn are responsible for the ferrimagnetic ordering.
(b) Primitive unit cell of \TMA.
In our model, only Ti1~(A), Ti2~(B), and Mn~(C) sublattices are taken into account.
(c)~Inter-sublattice nearest-neighbor hoppings.
(d)~Intra-sublattice nearest-neighbor hoppings.
Every sublattice has the same hopping vectors.
} 
\label{fig:lattice}
\end{figure}

 In this section, we introduce a simple tight-binding model of the compensated ferrimagnetic Weyl semimetal \TMA.
 The crystal structure of \TMA is shown in Fig.~\ref{fig:lattice}(a).
 Each sublattice (Ti1, Ti2, Mn, and Al) forms a face-centered-cubic (FCC) lattice, 
and thus the primitive unit cell is the FCC type [Fig.~\ref{fig:lattice}(b)].
 By focusing on pairs of the sublattices, Ti1-Al~(orange and gray) and Ti2-Mn~(red and blue) form the rocksalt structure.
 The rest of the combinations, e.g., Ti1-Ti2 or Ti1-Mn, form diamond or zincblende structures.

 Our model consists of 
a single orbital from each of Ti1~(A), Ti2~(B), and Mn~(C) atoms.
We neglect Al orbitals, 
which are not responsible for the magnetism, 
for simplicity.
We explain our model Hamiltonian $H$ by dividing into three components,
$H = H\_{t} + H\_{exc} + H\_{SOC}$,
where $H\_{t}$ represents the hopping,
$H\_{exc}$ the exchange coupling, and
$H\_{SOC}$ the spin-orbit coupling (SOC).

 The hopping component $H\_{t}$ reads
\begin{align}
 H\_{t} =  
       & -\!\sum_{\braket{ij} s}\!
            \(   t\_{AB} a^{\dagger}_{is} b_{js}
               + t\_{BC} b^{\dagger}_{is} c_{js}
               + t\_{CA} c^{\dagger}_{is} a_{js} + {\rm h.c.}\!
            \) \nonumber\\
       & -\!\sum_{\braket{ij} s}\!
            \(   t\_{AA} a^{\dagger}_{is} a_{js}
               + t\_{BB} b^{\dagger}_{is} b_{js}
               + t\_{CC} c^{\dagger}_{is} c_{js}
            \) \nonumber\\
       & +\!\sum_{i s}
            \(   \epsilon\_{A} a^{\dagger}_{is} a_{is}
               + \epsilon\_{B} b^{\dagger}_{is} b_{is}
               + \epsilon\_{C} c^{\dagger}_{is} c_{is}
            \),
 \label{eqn:hamiltonian}
\end{align}
where 
$a_{is}$, $b_{is}$, and $c_{is}$ are the annihilation operators
for electrons at Ti1 (A), Ti2 (B), and Mn (C) sites, respectively. 
 The first line corresponds to the inter-sublattice nearest-neighbor hopping [Fig.~\ref{fig:lattice}(c)].
 The second line corresponds to the intra-sublattice nearest-neighbor hopping [Fig.~\ref{fig:lattice}(d)].
 The third line is the on-site energy.

 The exchange component is 
\begin{align}
    H_{\rm exc} &= -\sum_{i} {\bm m} \cdot
                    \(   J\_{A} {\bm s}_{{\rm A},i} 
                       + J\_{B} {\bm s}_{{\rm B},i}
                       - J\_{C} {\bm s}_{{\rm C},i}
                    \),
\end{align}
%
where
${\bm s}_{{\rm A},i} = a^{\dagger}_{is} \({\bm \sigma}\)_{ss'} a_{is'}$ is the itinerant spin operator of A site, 
and the same applies to ${\bm s}\_{B}$ and ${\bm s}\_{C}$.
${\bm m}$ is the unit vector that is parallel to the magnetic moment of Ti (A and B) and antiparallel to that of Mn (C).
$J_\alpha\ (\alpha=\mathrm{A,B,C})$ are the coupling strength.

 The SOC $H\_{SOC}$ originates from the
broken inversion symmetry of the crystal structure.
%
%
%
 The dominant symmetry breaking comes from the imbalance between TI1 and Al sublattices.
 We assume the amplitudes of the SOC terms for the Ti2-Ti2 and Mn-Mn hoppings are the same, for simplicity.
 Since the atomic number of Ti and Mn are close to each other, compared with Al, we neglect the SOC for the Ti1-Ti1 hopping.
 The SOC term can be described in a Fu-Kane-Mele-like form \cite{Fu2007},
\begin{align}
 H\_{SOC} = -i\frac{8\lambda\_{SOC}}{\sqrt{2} a^2} 
          & \sum_{\braket{ij}} 
            \left[   b^{\dagger}_{is} (\bm{d}^{{\rm B}ij}_{1} \times \bm{d}^{{\rm B}ij}_{2}) 
                     \cdot (\bm{\sigma})_{ss'} b_{js'}
            \right. \nonumber\\
          & \left. +~c^{\dagger}_{is} (\bm{d}^{{\rm C}ij}_{1} \times \bm{d}^{{\rm C}ij}_{2})
                     \cdot (\bm{\sigma})_{ss'} c_{js'}
            \right].
 \label{eqn:Hsoc}
\end{align}
 Here $\lambda\_{SOC}$ is the strength of the SOC.
 $\bm{d}^{\alpha ij}_{1,2}$ are the two nearest-neighbor hopping vectors from the site $i$ to $j$ of the sublattice $\alpha$.
 Note that $\bm{d}^{{\rm B}ij}_{1,2} = {\bm n}_{l_{ij}}$
and $\bm{d}^{{\rm C}ij}_{1,2} = -{\bm n}_{l_{ij}}\ (l_{ij}=1,2,3,4)$ 
with 
${\bm n}_{1} = \frac{a}{4}(1,-1,-1)$,
${\bm n}_{2} = \frac{a}{4}(-1,1,-1)$,
${\bm n}_{3} = \frac{a}{4}(-1,-1,1)$, and
${\bm n}_{4} = \frac{a}{4}(1,1,1)$.

 We set the lattice constant $a=1$ for simplicity. 
 The hopping parameters are set to 
$t\_{AB}=1.1t_{0}$,
$t\_{BC}=0.4t_{0}$,
$t\_{CA}=1.2t_{0}$,
$t\_{AA}=0.05t_{0}$,
$t\_{BB}=0.85t_{0}$, and
$t\_{CC}=-0.05t_{0}$.
 On-site energies 
$\epsilon\_{A} = \epsilon\_{B} = \epsilon\_{C} = -2.15t_{0}$.
 The strengths of the exchange coupling 
$J\_{A} = J\_{B} = 0.7t_{0}$,
$J\_{C} = 1.7t_{0}$.
 The strength of the SOC 
$\lambda\_{SOC} = -0.2t_{0}$.
 We set the energy unit
$t_{0}= 0.33$ eV.
 The parameters are set so that the energy bands, density of states, and AHCs become consistent with first-principles calculations
as discussed later.

\section{Electronic structure}

\begin{figure*}[htbp]
   \centering
   \includegraphics[width=1.0\hsize]{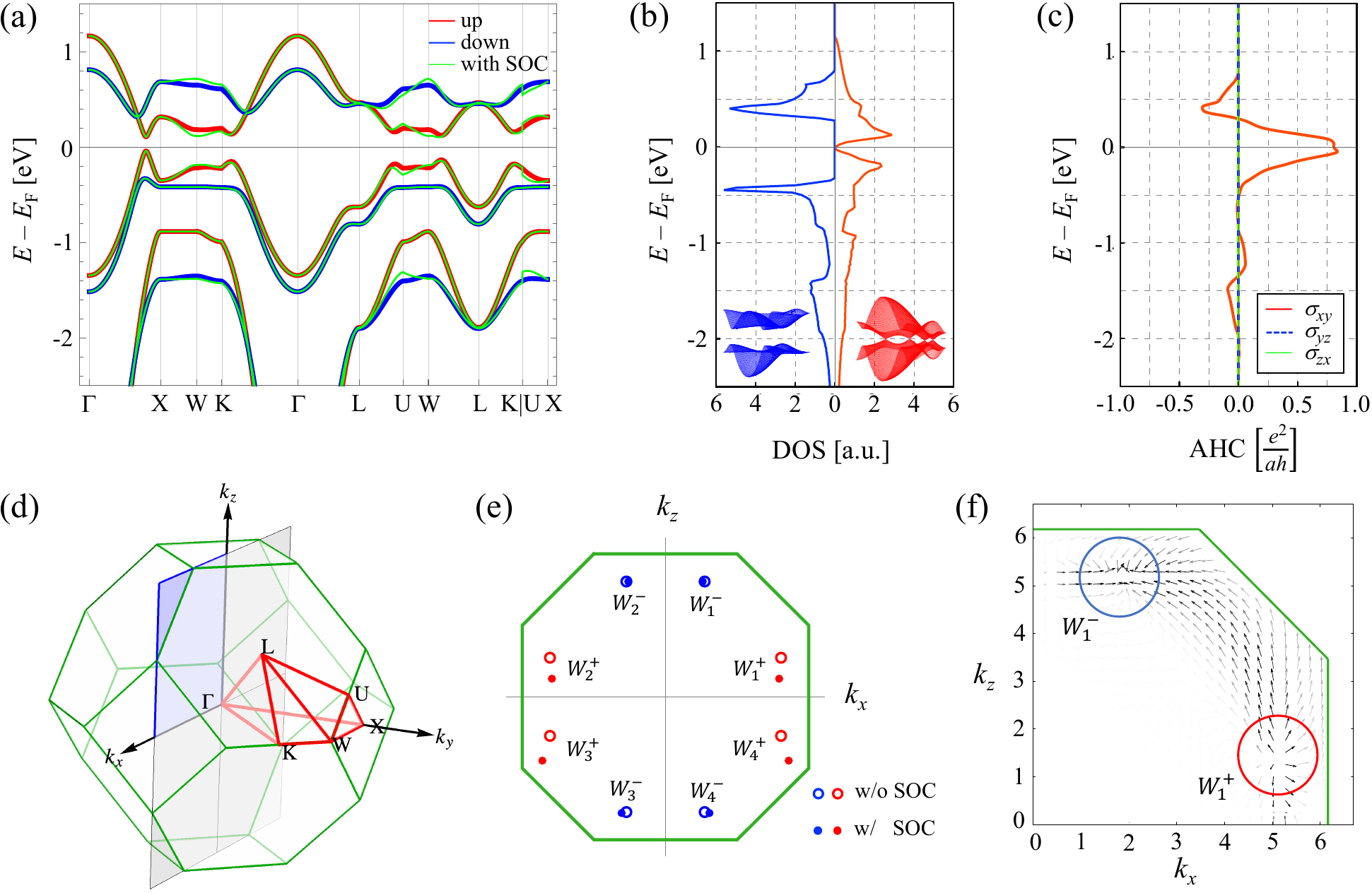}
\caption{
(a)~Band structure along the high symmetry lines.
(b)~The density of states for (red) up- and (blue) down-spin states and 
(c)~anomalous Hall conductivities $\sigma_{xy}$, $\sigma_{yx}$, $\sigma_{zx}$ as a function of energy.
(d)~The Brillouin zone and high symmetry lines of the system.
%
(e)~Configuration of the Weyl points with and without spin-orbit coupling and
(f)~the Berry curvature with spin-orbit coupling on the $k_{x}$-$k_{z}$ plane.
}
\label{fig:band}
\end{figure*}

Now, we study the electronic structure of this model.
The momentum representation of the Hamiltonian is explicitly expressed as,
\begin{widetext}
\begin{align}
\mathcal{H}(\bm{k})
    &=
    \begin{pmatrix}
         - t\_{AA} f({\bm k}) - \!J\_{A}{\bm m} \cdot {\bm \sigma}
          & -t\_{AB} g({\bm k})
          & -t\_{CA} g^{*}({\bm k})\\
        -t\_{AB} g^{*}({\bm k}) 
          & - t\_{BB} f({\bm k}) + \lambda\_{SOC} {\bm R}({\bm k}) \cdot {\bm \sigma}  
             -\!J\_{B}{\bm m} \cdot {\bm \sigma}
          & -t\_{BC} \sum_{i=x,y,z} \cos\frac{k_{i}}{2}\\
        -t\_{CA} g({\bm k}) 
          & -t\_{BC} \sum_{i=x,y,z} \cos\frac{k_{i}}{2} 
          & - t\_{CC} f({\bm k}) - \lambda\_{SOC} {\bm R}({\bm k}) \cdot {\bm \sigma}
            +\!J\_{C}{\bm m} \cdot {\bm \sigma}
    \end{pmatrix}\!.
\end{align}
\end{widetext}
%
Here, $f({\bm k})$ and $g({\bm k})$ correspond to 
the FCC and diamond hoppings, respectively, and are defined as,
  \begin{equation}
        f({\bm k}) \!=\! 4\! \(\!  \cos\!\frac{k_{x}}{2} \!\cos\!\frac{k_{y}}{2} 
                        \! + \!\cos\!\frac{k_{y}}{2} \!\cos\!\frac{k_{z}}{2} 
                        \! + \!\cos\!\frac{k_{z}}{2} \!\cos\!\frac{k_{x}}{2} \!  \)\!,
  \end{equation}
%
%
  \begin{equation}
        g({\bm k}) = \exp \(-i \sum^{4}_{\mu = 1} {\bm n}_{\mu} \cdot {\bm k} \).
  \end{equation}
${\bm R}({\bm k})$ corresponds to the FCC hopping with SOC and is defined as, 
\begin{align}
{\bm R}({\bm k})
        &=
        \begin{pmatrix}
            \sin\frac{k_{x}}{2} \left( \cos\frac{k_{y}}{2} - \cos\frac{k_{z}}{2} \right)\\
            \sin\frac{k_{y}}{2} \left( \cos\frac{k_{z}}{2} - \cos\frac{k_{x}}{2} \right)\\
            \sin\frac{k_{z}}{2} \left( \cos\frac{k_{x}}{2} - \cos\frac{k_{y}}{2} \right)
        \end{pmatrix}.
\end{align}
We use the magnetic moment pointing in the out-of-plane direction as ${\bm m}=(0,0,1)$.
By solving the eigenvalue equation $\mathcal{H}(\bm{k})|n, \bm{k}\rangle = E_{n\bm{k}}|n,\bm{k}\rangle$, 
the eigenvalues $E_{n\bm{k}}$ and eigenstates $|u_{n\bm{k}}\rangle$ are obtained.
Here, $n=1,2,...,6$ is the band index labeled from the bottom.
Figures~\ref{fig:band}(a) and~\ref{fig:band}(b) show the band structure along the high-symmetry lines and the density of states~(DOS) as a function of energy, respectively.
The high-symmetry lines are shown in Fig.~\ref{fig:band}(d).
%
%
We assume that the Fermi level $E_{\rm F}$ is the energy which is satisfied $4/6$ filling, and is being set as $E/t_0=0$.
At the energy $E_{\rm F}$ , we have the Weyl points as discussed later.
In Fig.~\ref{fig:band}(a),
red and blue lines indicate up and down spin band, respectively, in the absence of SOC.
Green lines indicate those in the presence of SOC.
Here, we focus on the spin up bands near $E/t_0=0.0$.
As shown in Fig.~\ref{fig:band}(b), the spin up bands show the local minimum of the DOS near the Fermi level~($E/t_0=0.0$).
This minimum is consistent with the result obtained by the first-principles calculations and may be significant to the large anomalous Hall angle~\cite{Shi2018}.
As shown in the insets of Fig.~\ref{fig:band}(b), the energy spectrum of these majority spin bands show the gapless linear dispersions, corresponding to the Weyl points, as discussed later.
On the other hand, down-spin state~(blue) is gapped at the Fermi level~($E/t_0=0.0$).

Next we study the Weyl points structure in Brillouin zone.
We show that our model describes the Weyl points structure similarly located to the result obtained by the first-principles calculations.
From our model Hamiltonian $\mathcal{H}(\bm{k})$ 24 gapless nodes~(Weyl points) are obtained between $n=4$ band and $n=5$ band.
The $k_x$-$k_z$ plane has the Weyl points as shown in Fig.~\ref{fig:band}(e).
Each Weyl points with chirality $+$~($-$) is labeled by $W^{+}_{\alpha}$~($W^{-}_{\alpha}$).
$\alpha$ distinguish the pairs of the Weyl points.
To characterize these nodes by chirality, we compute the Berry curvature $\bm{b}_{n \bm{k}} = \nabla_{\bm{k}} \times \bm{a}_{n \bm{k}}$ of the $n = 4$ band.
Here, $\bm{a}_{n \bm{k}}$ is the Berry connection, defined as $\bm{a}_{n \bm{k}} = -i \langle n,\bm{k} |\nabla_{\bm{k}}| n,\bm{k} \rangle$.
Figure~\ref{fig:band}(f) shows the Berry curvature distribution on $k_{x}$-$k_{z}$ plane.
The strength of the shade of the arrows is the amplitude of $\bm{b}(\bm{k})$.
The red and blue circles indicate the sources~($+$) and sinks~($-$) of the Berry curvature.
The configuration of the Weyl points is consistent with those obtained by the first-principles calculations~\cite{Shi2018}.

\section{Anomalous Hall effect}

Let us study the AHE in this section.
Using the Kubo formula~\cite{Nagaosa2010}, 
the anomalous Hall conductivities~(AHCs) $\sigma_{ij} \ (i \neq j)$ can be expressed as follows,
%
  \begin{equation}\label{eqn:AHC}
     \begin{split}
        \sigma_{ij} &= 
                     -\frac{i e^2}{\hbar} \int_{\rm BZ} \frac{d^{3} \bm{k}}{(2\pi)^3} 
                     \sum_{m \neq n} 
                     \frac{f(E_{n\bm{k}}) - f(E_{m\bm{k}})}
                          {E_{n\bm{k}} - E_{m\bm{k}} + i\eta}
                                              \\
                    &\ \ \ \ \ \ \ \ \ \ \ \ \ \ \ \ \ \ \ 
                     \times
                     \frac{\langle n,\bm{k} |\hbar v_{i}| m,\bm{k} \rangle \langle m,\bm{k} |\hbar v_{j}| n,\bm{k} \rangle}
                          {E_{n\bm{k}} - E_{m\bm{k}}}.
     \end{split}
   \end{equation}
Here, $f(E) = 1/(e^{\beta(E-E_{\rm F})}+1)$ is the Fermi-Dirac distribution function and $ \hbar \bm{v} = \frac{\partial \mathcal{H}(\bm{k})}{\partial \bm{k}}$ is the velocity operator.
Figure~\ref{fig:band}(c) shows $\sigma_{xy}$, $\sigma_{yz}$, and $\sigma_{zx}$ as a function of energy.
Near the energy of the Weyl points, the $\sigma_{xy}$ is maximized
, while $\sigma_{yz}$ and $\sigma_{zx}$ vanish.
The value of $\sigma_{xy}$ at the peak is $\sigma_{xy} \approx 0.80~[e^2/a h] = 496~{\rm S/cm}$, which reasonably agrees with the result obtained by first-principles calculations, $550~{\rm S/cm}$~\cite{Shi2018}.\\\indent
%
Then we study the relation between the AHCs and the configuration of the Weyl points.
As discussed in the introduction, when the Fermi level is located at the Weyl points,
the AHCs can be calculated with the distances between the Weyl points with opposite chirality $\Delta\bm{Q}^{\alpha}= ({\bm K}^{+}_{\alpha}-{\bm K}^{-}_{\alpha})$ \cite{Burkov2011},
  \begin{equation}\label{eqn:weyl}
   \begin{split}
        \sigma^{\rm Weyl}_{ij} &=\frac{e^2}{(2\pi)^2 \hbar} \sum_{\alpha=1}^{12}\sum_{k}\epsilon_{ijk} \Delta Q_{k}^{\alpha}.
     \end{split}
   \end{equation}
Here ${\bm K}^{+(-)}_{\alpha}$ is the position of the Weyl points with chirality~$+(-)$.
This $\sigma^{\rm Weyl}_{xy}$ is being $\approx 0.75 \ [{e^2/a h}]$, which is in good agreement with the maximized value obtained by the Kubo formula, shown in Fig.~\ref{fig:band}(c).
Therefore, the AHE in our model mainly originates from the Weyl points.
%
%


%
%
%
Next, we discuss the role of SOC for the Weyl points and the AHC.
In the absence of SOC, the Weyl points are
 distributed symmetrically, as represented by empty circles in Fig.~\ref{fig:band}(e).
This can be anticipated by the crystal symmetry of the system \cite{Shi2018}. 
In this case, the sum $\sum_{\alpha}\Delta\bm{Q}^{\alpha}$ cancels,
and thus AHC $\sigma^{\rm Weyl}_{xy}$ vanishes.
On the other hand, in the presence of SOC, the positions of the Weyl points are shifted as represented by filled circles in Fig.~\ref{fig:band}(e).
For the pairs ($W^{+}_{1}$,$W^{-}_{1}$) and ($W^{+}_{2}$,$W^{-}_{2}$), the distance between the Weyl points $\Delta Q^{\alpha=1,2}_z$ becomes longer.
Whereas, for the pairs of ($W^{+}_{3}$,$W^{-}_{3}$) and ($W^{+}_{4}$,$W^{-}_{4}$), $\Delta Q^{\alpha=3,4}_z$ becomes shorter.
Therefore, the cancellation of the sum $\sum_{\alpha}\Delta\bm{Q}^{\alpha}$ is broken, giving rise to the finite AHC $\sigma^{\rm Weyl}_{xy}$.
We note that the Weyl points in other planes are shifted in a similar manner to those in the $k_x$-$k_z$ plane.

\begin{figure}[t]
   \centering
   \includegraphics[width=1.0\hsize]{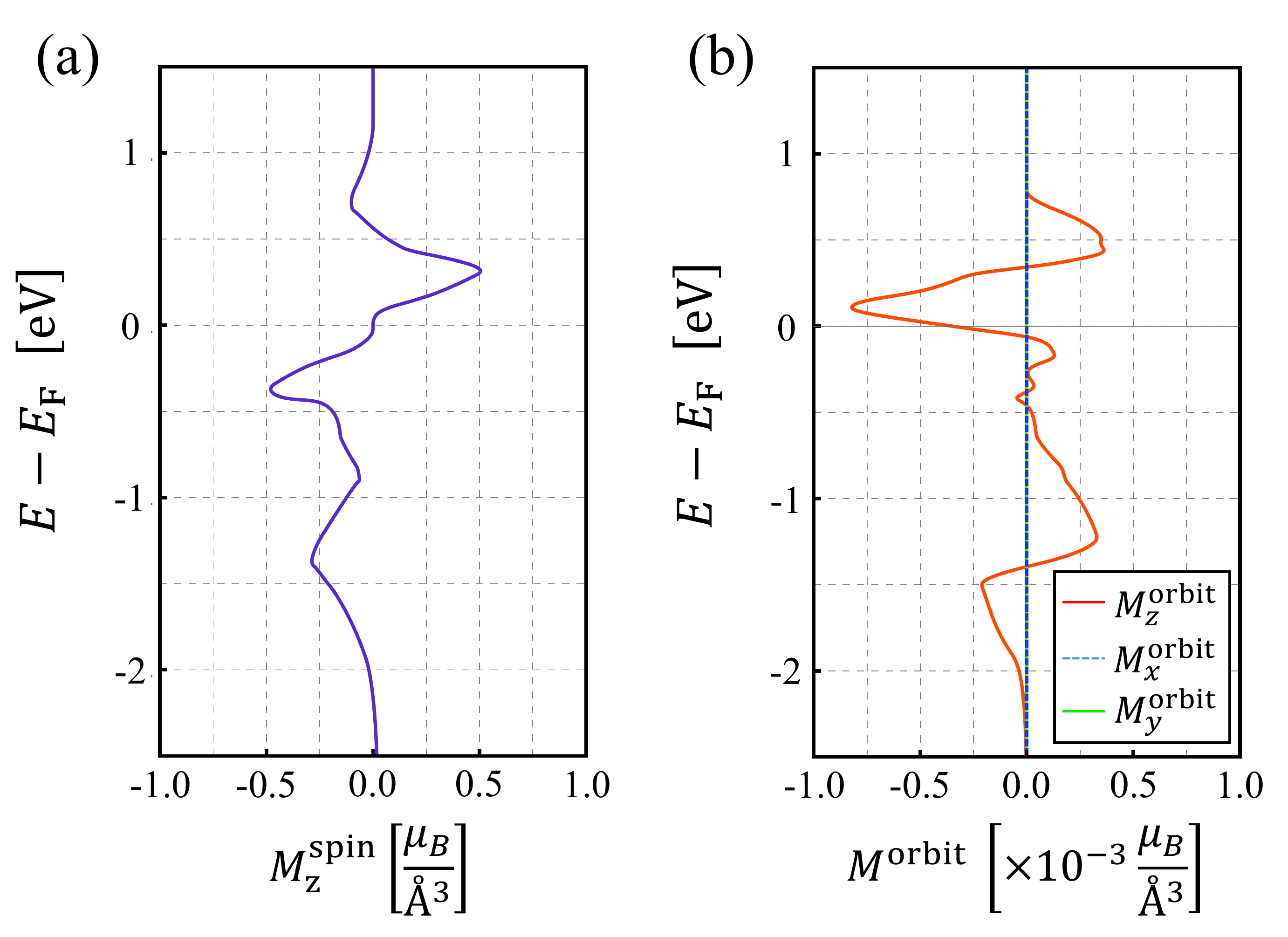}
\caption{
(a) Spin magnetization $M^{\rm spin}_z$
and
(b) each component of orbital magnetization ${\bm M}^{\rm orbit}$ as a function of energy 
for the magnetic moment $\bm{m}$ parallel
 to the $z$ axis.
} 
\label{fig:mag}
\end{figure}

\section{Spin and Orbital magnetizations}
In this section, we study spin and orbital magnetizations in our model.
The spin magnetization is calculated by the following equation,
  \begin{equation}\label{eqn:spinmag}
     \begin{split}
        M^{\rm spin}_z &= \mu_{\rm B} \int^{E_{\rm F}}_{E_0} [D_{\uparrow}(\epsilon) - D_{\downarrow}(\epsilon)] d\epsilon.
     \end{split}
   \end{equation}
Here, $\mu_{\rm B}$ is the Bohr magneton, and $E_0$ is the lower band edge.
$D_\uparrow(\epsilon)$ [$D_\downarrow(\epsilon)$] is the DOS for the up-spin~(down-spin).
The spin magnetization as a function of energy is shown in Fig.~\ref{fig:mag}(a).
Recall that the magnetic moments of Ti and Mn are parallel to the $z$ axis.
$M^{\rm spin}_z$ vanishes at $E=E_{\rm F}$, indicating the compensated ferrimagnetic phase.
Owing to this characteristic DOS, finite spin magnetization may be obtained by tuning $E_{\rm F}$, 
using a gate voltage, for instance.
When $E_{\rm F}$ is increased~(decreased), 
finite and positive~(negative) spin magnetization might be generated.
This indicates that, in \TMA, one can induce and switch the spin magnetization electrically.
Recently, the electrical control of ferrimagnetic systems has been studied from viewpoint of a 
functional magnetic memory, for example in Ref.~\onlinecite{Huang2021}.



%
%
%
%
%
%

Next we study the orbital magnetization~\cite{Ceresoli2006,Shi2007, Ominato2019},
  \begin{equation}\label{eqn:orbitmag}
     \begin{split}
        M^{\rm orbit}_k &= \frac{e}{2\hbar} \int_{\rm BZ} \frac{d^{3} \bm{k}}{(2\pi)^3} \sum_{i,j} \epsilon_{ijk} \\
        &{\rm Im} \sum_{m \neq n} f(E_{n\bm{k}})
        \frac{\langle n,\bm{k} |\hbar v_{i}| m,\bm{k} \rangle 
        \langle m,\bm{k} |\hbar v_{j}| n,\bm{k} \rangle}{(E_{m\bm{k}} - E_{n\bm{k}})^2} \\
        &\ \ \ \ \ \ \ \ \ \ \ \ \ \ \ \ \ \ \ \ \ \ \ \ \ \ 
        {\times}(E_{n\bm{k}} + E_{m\bm{k}} - 2 E_{\rm F}).
     \end{split}
   \end{equation}
Figure~\ref{fig:mag}(b) shows each component of the orbital magnetization ${\bm M}^{\rm orbit}$ as a function of energy.
We find that $M^{\rm orbit}_{z}$ can be finite,
while $M^{\rm orbit}_{x}$ and $M^{\rm orbit}_{y}$ vanish.
Although orbital magnetization is usually much smaller than spin magnetization,
$M^{\rm orbit}_{z}$ is finite at $E\_F$, where $M^{\rm spin}_{z}$ is fully compensated.
As discussed later, this small but finite orbital magnetization might be used to switch the directions of magnetic moments on Ti and Mn sites.

\section{Magnetic Isotropy}

\begin{figure}[htbp]
   \centering
   \includegraphics[width=1.0\hsize]{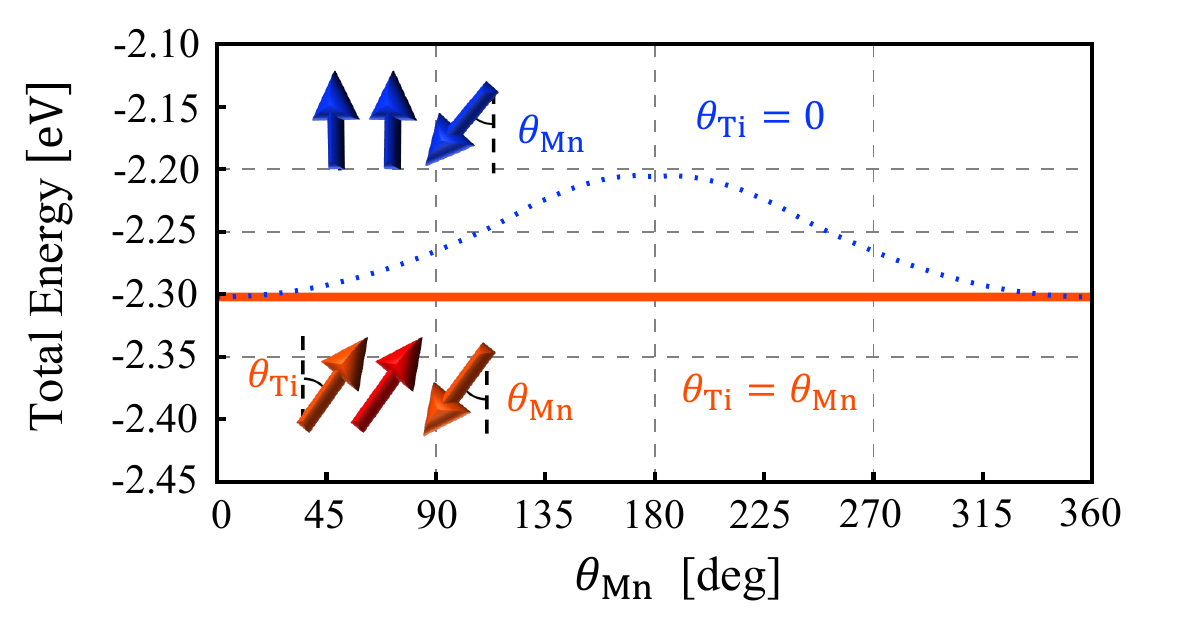}
\caption{
The magnetization-angle dependence of the total energy of electron, Eq.~\eqn{Ee}, 
when one 
rotates the magnetizations of Mn and Ti uniformly~(solid red line)
or the magnetization of Mn solely~(dotted blue line).
} 
\label{fig:anisoE}
\end{figure}

In this section, we study the magnetic anisotropy and angular dependences of the AHCs and orbital magnetization.
We start to study magnetic anisotropy.
Figure~\ref{fig:anisoE} shows the total energy of electrons, computed as  
\begin{align}
  E\_{e} &= \frac{1}{N_{\bm{k}}} \sum_{n=1}^{6} \sum_{{\bm k}} E_{n{\bm k}} f(E_{n{\bm k}}),
\label{eqn:Ee}
\end{align}
where $N_{\bm{k}}$ is the number of $\bm{k}$-mesh.
%
Here, we consider the two tilted configurations in which the magnetic moments of (a) Mn only or (b) both Mn and Ti are tilted in $x$-$z$ plane.
%
Their tilting angles are denoted by the angles (a)~$\theta_{\rm Mn}$ and (b)~$\theta_{\rm Mn}=\theta_{\rm Ti}$.
The schematic figures for these situations are shown in the insets of Fig.~\ref{fig:anisoE}.
In case (a), as the blue dotted line indicates, the electron energy is maximized at $\theta_{\rm Mn}=180^{\circ}$, indicating the stability of the compensated ferrimagnetic ordering.
On the other hand, in case (b), the electron energy is independent of $\theta_{\rm Mn}=\theta_{\rm Ti}$ as the red solid line indicates.
Therefore, in our model, the compensated ferrimagnetic ordering does not show the easy-axis magnetic anisotropy.
%
With these results, the ferrimagnetic interaction between Ti and Mn can be estimated as $0.1t_{0} \approx 33~{\rm meV}$, when $t_0$ = 0.33 eV is assumed.

%

%
\begin{figure}[htbp]
   \centering
   \includegraphics[width=1.0\hsize]{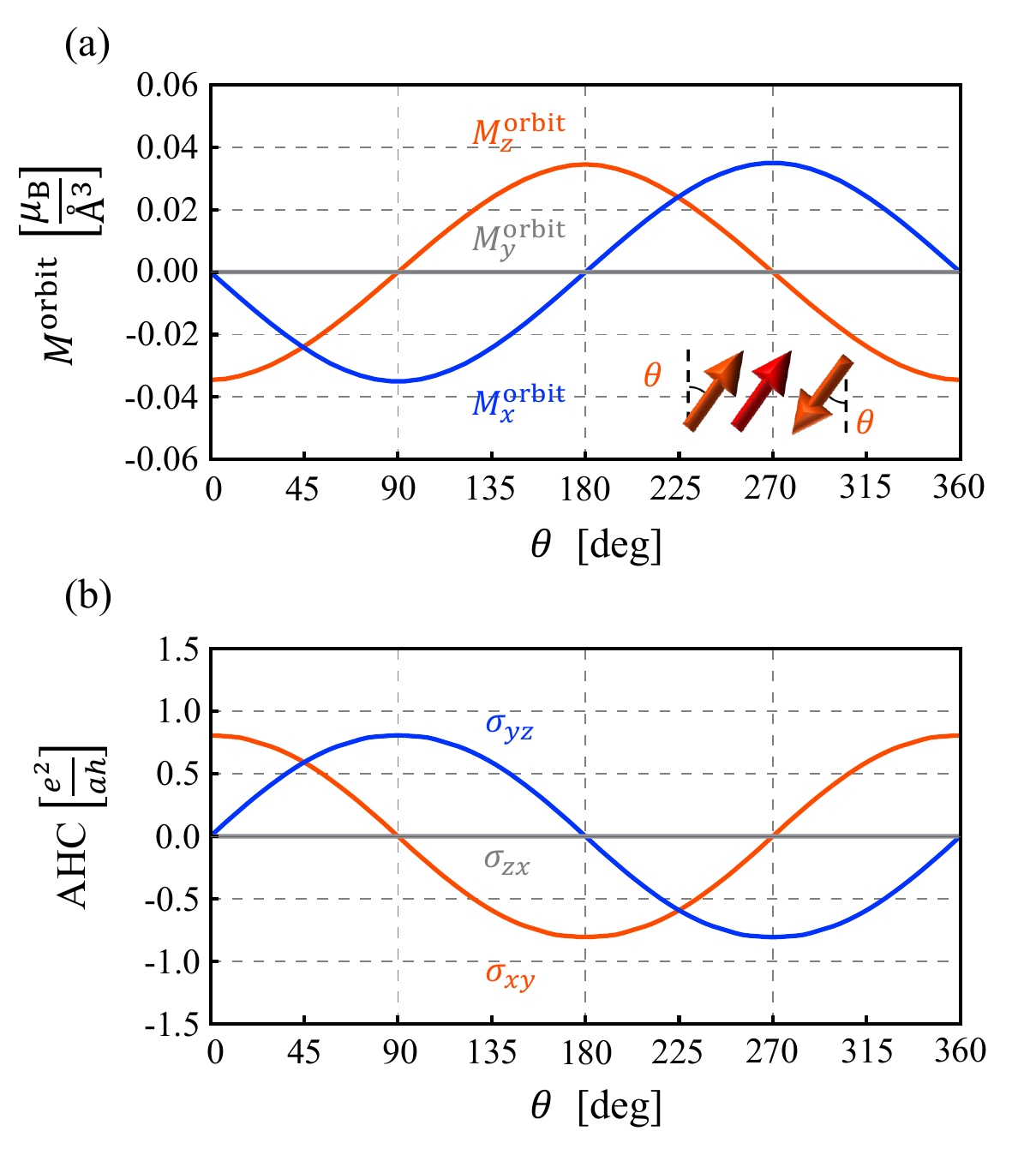}
\caption{
The magnetization-angle dependence of
(a) orbital magnetization and 
(b) AHCs 
at $E=E_{\rm F}$ 
when one changes the direction of magnetic moment uniformly
in the $x$-$z$ plane.
}
\label{fig:aniso}
\end{figure}

Then, we study a correlation between magnetic moments and orbital magnetization.
In the previous section, we showed that, at the energy of the Weyl points, our model shows compensated spin magnetization and finite orbital magnetization.
%
In the following, we tilt the magnetic moments on both Ti and Mn sites in the $x$-$z$ plane, as shown in the inset of Fig.~\ref{fig:aniso}(a).
Figure~\ref{fig:aniso}(a) shows
each component of the orbital magnetization as a function of $\theta$.
We find that the orbital magnetization follows the direction of the magnetic moments.
This feature can be used for the control of the magnetic moments, as similarly discussed in Refs.~\onlinecite{Ito2017,Ozawa2022}.
%
%
Under an external magnetic field, 
magnetic moments on each site and orbital magnetization couple with the field via Zeeman interaction.
However, as discussed in the previous paragraph, 
the ferrimagnetic coupling between the magnetic moments on Ti and Mn is stronger than 
the typical strength of the Zeeman interaction~($\approx 0.15~{\rm meV}$ at 1 T).
Thus, the Zeeman interaction of the magnetic moments on both sites cancels each other.
This cancellation implies that only the orbital magnetization couples with the external magnetic field.
In addition, owing to SOC, 
the directions of the magnetic moments are locked 
with the orbital magnetization, as shown in Fig.~\ref{fig:aniso}(a).
Consequently, 
the direction of the magnetic moments can be controlled by an external magnetic field, even without net magnetization.
%
%
%
%
%
%
%
%
%
%
The changes in the direction of the magnetic moments may be 
probed
 by the AHE.
Figure~\ref{fig:aniso}(b) shows the $\theta$
dependence
 of the AHCs.
We find the relation \
$\sigma_{ij} \propto -\sum_{k}\epsilon_{ijk} {M}^{\rm orbit}_{k}$.
This relation indicates that the directions of the magnetic moments are experimentally determined by measuring the AHCs.


\section{Conclusion}

In this paper, we constructed an effective model of compensated ferrimagnetic Weyl semimetal \TMA.
The Weyl points configuration in our model reasonably agrees with those obtained by the first-principles calculations.
The finite AHC in the presence of SOC can be understood by the shifting of the Weyl points.
At the energy of the Weyl points, the orbital magnetization is finite while the total spin magnetization vanishes.
The magnetic moments at each site are correlated with the orbital magnetization,
and can be controlled by the external magnetic field.



\acknowledgments

The authors would like to appreciate 
Y.~Araki and A.~Tsukazaki 
for valuable discussions.
This work was supported by
JST CREST, Grant No.~JPMJCR18T2
and by
JSPS KAKENHI, Grant Nos.~%
JP20H01830  
and
JP22K03446. 
A.~O.~was supported by
GP-Spin at Tohoku University. 


\bibliography{ref_Ti2}
\bibliographystyle{apsrev}

\end{document}